\documentstyle[11pt,aasms,tighten,flushrt]{article}
\begin{document}
%\normalsize
%\small

%\setcounter{page}{1}
% \noindent hbron1.tex:  Sept. 6, 2000.

\title{
CLASSIFICATION OF PLANETARY NEBULAE BY THEIR DEPARTURE FROM AXISYMMETRY}
%\vspace*{2.0cm}

\author{ 
Noam Soker and Ron Hadar$^1$}
\affil{
Department of Physics, University of Haifa at Oranim\\
%Mathematics-Physics\\
Oranim, Tivon 36006, ISRAEL \\
soker@physics.technion.ac.il \\      
$^1$ Permanent position: West Valley High School, Kibbutz Yifat, 30069,
Israel}

%\clearpage 
\begin{center}
\bf ABSTRACT
\end{center}

  We propose a scheme to classify planetary nebulae (PNe) according to
their departure from axisymmetric structure.  We consider only
departure along and near the equatorial plane, i.e., between the two
sides perpendicular to the symmetry axis of the nebula.
 We consider 6 types of departure from axisymmetry:
(1) PNe whose central star is not at the center of the nebula;
(2) PNe having one side brighter than the other;
(3) PNe having unequal size or shape of the two sides;
(4) PNe whose symmetry axis is bent, e.g., the two lobes in bipolar
PNe are bent toward the same side;
(5) PNe whose main departure from axisymmetry is in the
outer regions, e.g., an outer arc;
(6) PNe which show no departure from axisymmetry, i.e., any departure,
if it exists, is on scales smaller than the scale of blobs, filaments,
and other irregularities in the nebula.
 PNe which possess more than one type of departure are classified by
the most prominant type.
  We discuss the connection between departure types and
the physical mechanisms that may cause them, mainly due to
the influence by a stellar binary companion.
 We find that $\sim 50 \%$ of all PNe possess large-scale
departure from axisymmetry.
 This number is larger than that expected from the influence of
binary companions, namely $\sim 25-30 \%$.
We argue that this discrepancy comes from many PNe whose departure
from axisymmetry, mainly unequal size, shape, or intensity, results
from the presence of long-lived and large, hot or cool, spots on
the surface of their AGB progenitors.
 Such spots locally enhance mass loss rate, leading to a
deparure from axisymmetry, mainly near the equator, in the
descendant PN.
%====================================================================
%====================================================================
\bigskip


\clearpage

% ======================================================================
\section{INTRODUCTION}
% ======================================================================

 The rich varieties of planetary nebula (PNe) structures have
attracted considerable attention in recent years
(see Kastner, Soker \& Rappaport 2000 for a collection of papers with
additional references).
 The PNe were classified according to their large structure
(Schwarz, Corradi, \& Stanghellini 1992; Manchado {\it et al.}
1996, 2000; Stanghellini {\it et al.} 1999), with further divisions
according to microstructures, such as different types of blobs
(e.g., Goncalves, Corradi \& Mampaso 2001). 
 The most common classification is into Bipolar
(also called ``bilobal'' and ``butterfly'') PNe, which
are defined as axially symmetric PNe having two lobes with
an `equatorial' waist between them, and elliptical PNe, which 
have a large scale elliptically shaped shell, with no, or only small, 
lobes or waists.

 Most of these classifications consider axisymmetrical structures,
with little attention paid to the departure of PNe from axisymmetry.
PNe with departure from axisymmetry are defined as having substructures
which have no symmetry axes, or the symmetry axes of different
substructures either lack a common intersection point or the
illuminating star is displaced from the intersection point.
 By departure from axisymmetry we refer only to large-scale structures,
and not to small blobs, filaments, bubbles, etc.
 Note that point-symmetric PNe also depart from pure
axisymmetry, but we do not define them as having departure
if they can be built by pure rotations of the different symmetry axes
of the different substructures.
 Only if displacements of one or more of the symmetry axes
relative to the central star is detected, and/or large scale asymmetry
exists, would we define a point-symmetric PN as having departure
from axisymmetry.
 Small numbers of observational papers in recent years (e.g., Sahai 2000b;
Miranda {\it et al.} 2001b; Miranda, Guerrero, \& Torrelles 2001a)
consider the departure from axisymmetry of specific PNe
(hereafter termed ``departure'').
 The only process which causes departure and which attracted more
attention of observers was the interaction of PNe with the ISM
(e.g., Jacoby 1981;  Tweedy \& Kwitter 1996;
Kerber {\it et al.} 2000; Muthu, Anandarao \& Pottasch 2000;
Rauch {\it et al.} 2000).
 This process also stimulated more theoretical work (e.g.,
Borkowski, Sarazin, \& Soker 1990; Soker, Borkowski, \& Sarazin 1991; 
Villaver, Manchado, \& Garcia-Segura 2000; 
see Dgani 2000 for a recent review).
 This process is most likely to cause departure in the outskirts of
the PN.

 Binary star system progenitors of PNe may also lead to departure
from axisymmetry, mainly in the inner regions of PNe.
Both close binary companions in eccentric orbits
(Soker, Rappaport, \& Harpaz 1998) and
wide binary companions (Soker 1994; 1999) can lead to departure
(for a summary of processes with further references see section 1
of Soker \& Rappaport 2001, hereafter SR01). 
  SR01 conduct population synthesis to estimate the
number of PNe expected to possess detectable departure
as a result of a binary companion influence on the AGB progenitor wind.
 They also study the distribution of systems according to whether the
companion is a main-sequence star or a white dwarf companion
(the remnant of the initially more massive star).
 They find that  $\sim25 \%$ of all elliptical or circular PNe
and $\sim30-50 \%$ of all bipolar PNe are expected to
possess detectable departure from axisymmetry.
 Since $\sim10-15 \%$ of all PNe are bipolar, and the
rest are elliptical or circular,  they argue that
$\sim 27 \%$ of all PNe are expected to possess significant departure
from axisymmetric structure.
 SR01 examine a limited sample of PNe and argue for a satisfactory
agreement between the fraction of PNe possessing departure from axisymmetry
found in the population synthesis and their estimate from the
sample of PNe.
 SR01 consider an observed PN to possess departure
if the illuminating star is not at the center of the nebula, if one side
of the nebula is being brighter or more extended than the other,
if the two sides have different magnitude Doppler shifts, or
if there is any combination of these.
 However, they do not classify the PNe according to the different
types of departure.

 Our goal is to build a classification scheme that (1) can
easily be used by people studying specific PNe to classify them,
and (2) has some connection with physical processes that can cause
departure from axisymmetry.
 In section 2 we propose a classification scheme, and give typical
PNe for each class.
 Section 3 contains discussion and summary, including comparison
with theoretical expectations, mainly according to the results
of SR01.

% ======================================================================
\section{THE PROPOSED CLASSIFICATION}
% ======================================================================

 Our proposed classification for departure from axisymmetry
is summarized in the first column of Table 1.
 The classification is based on 5 basic types of departure and
a class of PNe without departure, or with a very small one that can't
be distinguished from small-scale irregularities.
 For each class we give a few typical examples of elliptical
(second column) and bipolar (third column) PNe which possess that
type of departure.
 In many cases a PN can possess more that one type of departure.
 For example, the two sides along the equatorial plane can show 
unequal size as well as unequal intensity.
 Each PN is classified according to the most prominent type of departure
from axisymmetry, according to our, sometimes subjective,
judgment.
  The sources for the typical PNe given in Table 1 are listed there.
The HST images can be found in the home page of the STScI, and
in a list compiled by  Terzian \& Hajian (2000;
http://ad.usno.navy.mil/pne/gallery.html); many of the the
different HST images are taken from the following papers:
Balick (2000), Balick {\it et al.} (1998), Bobrowsky {\it et al.} (1998),
Bond (2000), Borkowski, Blondin, \& Harrington (1997),
Corradi {\it et al.} (2000), Kwok, Su, \& Hrivnak (1998), Sahai (2000a,b),
Sahai \& Trauger (1998), Sahai {\it et al.} (1998a,b)
Su {\it et al.} (1998), and Terzian \& Hajian (2000).
 We now elaborate on the different classes, leaving the
discussion of the physical mechanisms for departure to
the next section.
 In the present paper we only discuss departure from axisymmetry
along the equatorial plane.
 This avoids cases where the appearance of departure, e.g.,
between one lobe and another in bipolar PNe, is simply an effect of
obscuration, e.g., by dust, because of inclination effects.
 Therefore, when writing `two sides' we refer to two sides of the
symmetry axis, i.e., on or near the equatorial plane, but not along
the symmetry axis.
 
The proposed classes are as follows.
\newline
\noindent {\it Central star not in the center:}
 These PNe possess a general axisymmetric (or point-symmetric)
structure, possibly beside another type of departure as listed below,
but the central star is not along the symmetry axis
(or the intersection of a few axes in point-symmetric PNe).
\newline
\noindent {\it Unequal intensity:} 
 These PNe possess a general axisymmetric (or point-symmetric)
structure, but the intensity over a large area on one side is
higher than that on the other side.
 In the present paper we refer mostly (but not solely; see the
discussion in the next paragraph) to unequal intensity in the optical
band.
\newline
\noindent {\it Unequal size or shape:}
 In these PNe the most prominent departure is unequal size or shape of
the two sides of the symmetry axis. 
 Here, as with PNe with unequal intensity, the departure can reveal
itself in some bands but not in others (SR01).
 For example, the general structure of the Red Rectangle (BD-10-1476),
which contains a close eccentric binary system with $P_{\rm orb} = 318$
days and $e=0.38$ (Waters {\it et al.} 1998), 
is highly axisymmetric, up to a distance of $\sim1 ^\prime$ from the
central star (e.g., Van Winckel 2000).
However, the 10 $\mu$m map presented by Waters {\it et al.}
(1998; their fig. 3) shows a clear departure from axisymmetry on scales
of $\sim5 ^{\prime \prime}$ from the central star.
 Their contour map shows that the equatorial matter is more
extended on the west side.
 Another example is the Egg nebula which appears axisymmetric in the
HST optical image, while only the NICMOS image shows a clear departure
from axisymmetry (Sahai {\it et al.} 1998a,b).
\newline
\noindent {\it Bent:}
 These PNe contain an elongated structure near the symmetry axis,
e.g., two lobes or two jets, one at each side of the equatorial
plane, which are bent to the same side, i.e., there is a mirror
symmetry about the equatorial plane 
(not to be confused with point-symmetry, where the two structures
are bent in opposite directions).
 The bipolar PN NGC 2899 (Corradi \& Schwarz 1995) has two lobes bent
to the same side, while the elliptical PNe NGC 3242 and NGC 6826
(Balick 1987; see also HST images) contain two FLIERS (or ansae)
bent to the same side. 
\newline
\noindent {\it Departure in outer regions:}
 When a PN possesses departure in its very outer region
it belongs to this class.
 The departure can be in the shape, size or intensity of the two sides;
in most cases all three exist, hinting at an interaction of the PN
with the ISM.
 Many examples are given in Tweedy \& Kwitter (1996).
 Note that many PNe are old and large and it is hard
to tell whether they are intrinsically bipolar or elliptical
(or round) PNe, and whether they had departure not caused by the ISM;
we classified them as elliptical PNe.
 HDW 5 (PNG 218.9-10.7), for example, is classified by Manchado
{\it et al.} (1996) as a bipolar PN.
 However, it seems that the structure is more complicated, and it is hard
to tell whether this PN is really a bipolar PN.
\newline
\noindent {\it Very small (no) departure.}
 By very small departure we refer to PNe where we could not identify any
large-scale departure from axisymmetry (or from point-symmetry).
 Any departure, if it exists, is smaller than the typical size of blobs,
filaments, loops, and other small-scale irregularities in the nebula.
 A good example is NGC 6543, a very complicate point-symmetric PN.
Despite the many loops, filaments, and blobs, the PN possesses no
departure from point-symmetry on a scale larger than the typical
size of these irregularities.

% ======================================================================
\section{DISCUSSION AND SUMMARY}
% ======================================================================

  A discussion of physical processes that can cause departure from
axisymmetry is in SR01, where more details can be found.
Here we mention briefly the possible connection between these processes
and the classification scheme proposed in the previous section. 
 Four main processes can result in large-scale deviations from
axisymmetry.
\newline
{\it 1. Interaction with the ISM.}
In this case the most prominent features are on the outskirts of the
nebula (e.g., Tweedy \& Kwitter 1994, 1996; Rauch {\it et al.} 2000).
 Departures in the inner regions may still result from binary progenitors.
\newline
{\it 2. Local mass loss events.}
 If one or a few long-lived hot or cool spots exist on the surface
of the AGB star during the mass-loss process itself, they can lead
to enhanced mass-loss rates in one or a few particular directions.
 This process seems to be important in massive stars
(e.g., as suggested for the $\sim30 M_\odot$ star HD 179821
by Jura \& Werner 1999).
The question is whether this process can operate efficiently in
AGB stars, where strong convection may not allow such spots to
live long enough.
 Our results (see below) indicate that this process may occur
in AGB stars as well.
 Such large spots in the equatorial plane will cause the descendant
PN to have the `unequal size or shape' or `unequal intensity'
departure type.
 The presence of stochastic mass-loss process from the AGB star
and instabilities in the outflowing gas are the largest sources
of uncertainty when comparing theory and observations.
\newline
{\it 3. A wide binary companion.}
 This applies to the case where the AGB star has a wide binary companion
(Soker 1994). Here the departure from axisymmetry results simply from
the fact that the mass loss from the AGB star occurs while it is moving
in its orbit.
Interesting effects due to the orbital motion occur for a wide range
of orbital periods (Soker 1994; Mastrodemos \& Morris 1999; SR01).
 This type of systems explains many elliptical PNe belonging to
the `unequal size or shape', `unequal intensity', or `bent'
types.
\newline
{\it 4. A close binary companion in an eccentric orbit.}
 This occurs when the companion is close enough to influence the
mass-loss process from the AGB star and/or from the system as a whole,
and the eccentricity is substantial (Soker, Rappaport, \& Harpaz 1998).
 If the companion is close enough to significantly influence the
mass-loss process then most likely a bipolar PN is formed.
 This process is responsible for many bipolar PNe belonging to
`unequal size or shape', `unequal intensity', or `bent' types
of departure.

 Unequal intensity may result from collisions on one side of the nebula
of two segments of the wind blown at different orbital phases,
e.g., a spiral structure (Soker 1994; Mastrodemos \& Morris 1999).
 `Bent' PNe are those with departure along the symmetry (polar) axis.
 The flow along the polar directions is faster than that along
the equatorial plane. Hence in `bent' PNe the orbital
velocity of the mass-losing AGB star is relatively high, i.e.,
relatively close binary and/or massive companions.
 This explains why many bipolar PNe show departure in the slowly
expanding equatorial region (mostly unequal size and shape or
uneqaul intensity), but not along the fast-moving polar lobes.
 A quantitative study of these processes requires 3D gasdynamical
 numerical simulations. 

 SR01 find from their population synthesis study that $\sim25 \%$ of
all elliptical or circular PNe, and $\sim30-50 \%$ of all bipolar PNe,
are expected to possess detectable departure from axisymmetry.
 They also find that in the list of HST images compiled by
Terzian \& Hajian (2000; the sources to this list were cited in the
previous section) $\sim 30-60\%$ of all elliptical (and round) PNe,
and $\sim 35-60 \%$ of all bipolar PNe have some type of departure.
 We examine two large samples of high quality PNe images, 
the HST list and {\it The IAC Morphological Catalog of Northern
Galactic Planetary Nebulae} (Manchado {\it et al.} 1996).
 We find 69 good images in the HST list, with 44 elliptical (or round)
and 25 bipolar, and 179 PNe, of which 142 are elliptical
and 37 bipolar, in the IAC Catalog.
 There are 4 PNe common to both samples.
 We list the fraction (in percentage) of PNe belonging to
each type of departure separately for elliptical, columns 4-5,
and bipolar PNe, columns 6-7.
For example, there are 6 elliptical PNe with displaced central
star in the HST list, and a total of 44 elliptical in that list.
Hence we put a value of $6/44=14 \%$ for elliptical PNe with
``central star not at the center''.
 We do not include in this statistics PNe with departure in
outer regions only; these are usually large and may contain
another type of departure in the inner regions. 
 
 There are several sources of uncertainties in addition to the
statistical errors due to the small number of objects in each group.
($i$) There are difficulties in identifying the departure and its
type in many PNe.
($ii$) A selection exists due to the obscuration of the central star in
bipolar PNe.  This is the reason for the low number of bipolar PNe
with displaced central star.
($iii$) In comparing with theoretical expectations from binary
interaction, blobs, filaments, and other irregularities, caused
by stochastic mass-loss process or instabilities, may be mistaken
for large-scale departure caused by a binary companion.
  SR01 estimate that $\sim 20-25 \%$ bipolar PNe have large
blobs, filaments, and other irregularities.
 In the present study most of these PNe enter the
`unequal size or shape' or `unequal intensity' departure types.

 The fraction of elliptical PNe with departure found in the
present work is in the range of $46-52 \%$ (IAC Catalog and
HST list, respectively).
 This is compatible with the finding of SR01 from their analysis
of PN images, but larger than the fraction of $25 \%$
expected from binary interaction according to the population
synthesis (SR01).
 The fraction of bipolar PNe possessing departure is in the
range of $56-76 \%$ (HST list and IAC Catalog, respectively).
 The value from the HST list is within the range found
from analyzing the same data by SR01, while the value from
the IAC catalog is larger than these numbers.
 The fraction from the images analysis is much larger than the
theoretical expected fraction, $\sim 30-50\%$, from binary
interaction according to SR01.
 However, SR01 only study binary systems which avoided RLOF and/or
common envelope phases, which add up to only half of all progenitors
of bipolar PNe.
Therefore, it is quite possible that a large fraction
of bipolar PNe progenitors which went through a RLOF and/or
common envelope phase do acquire departure from axisymmetry from the
binary interaction.
An example is the PN NGC 2346 which have a central binary system
(Bond \& Livio 1990), and which possess departure from axisymmetry.
 Bipolar PN progenitors that were not studied by SR01 increase
the number of bipolar PNe expected to possess departure,
but still cannot overcome the large discrepancy.
 In addition, a large discrepancy exists in elliptical PNe also.
 The discrepancy between the number of PNe expected to acquire
departure from binary interaction and the number deduced from the
large sample of PNe suggests that long-lived large spots,
or local groupings of small spots, whether hot (Schwarzschild 1975)
or cool (Soker 1998), exist on the surface of AGB stars.
 These spots locally enhance, especially near the equatorial plane
(Soker 1998), the mass-loss rate, leading to a large-scale
departure from axisymmetry.
 Such a process was suggested by Jura \& Werener (1999) to have
occurred on the surface of the $\sim30 M_\odot$ star HD 179821,
and may also account for the asymmetry of Mira
(Karovska {\it et al.} 1997).
 Taking the different uncertainties into account, we crudely estimate
that $\gtrsim 15 \%$ of very late AGB stars have large cool (or hot)
spots when they reach the upper AGB, such that the enhanced mass-loss
rate above these, relatively long-lived, spots causes
the descendant PN to acquire departure from axisymmetry.
 The fraction can be much larger than $15 \%$, if many AGB stars
have a nonaxisymmetric mass-loss process due to a companion,
and in addition possess large spots on their surface.

 The main points of the present paper are as follows.
\newline
 (1) We proposed a classification scheme (first column of Table 1)
to the different types of departures from axisymmetry observed
in PNe.
\newline
 (2) We find that $\sim 50 \%$ of all PNe possess large-scale
departure from axisymmetry.
 The distribution among the different types (column 4-7 of
Table 1) is highly uncertain, for reasons discussed in the paper.
\newline
 (3) The number of PNe found to have departure from axisymmetry
is much larger than that expected from the influence of binary companions.
 This, we claim, results from the presence of long-lived and large,
hot or cool, spots on the surface of AGB stars, particularly near the
stellar equator, which locally enhance mass-loss rate, causing
the descendant PN to possess large-scale departure from axisymmetry.
 PNe that acquire their departure from axisymmetry
from such spots belong either to the `unequal size or shape'
type or the `unequal intensity' type. 
 We crudely estimate that $\gtrsim 15 \%$ of all stars on
the final phase of the AGB possess such spots. 
%====================================================================
%====================================================================
\bigskip

{\bf ACKNOWLEDGMENTS:}
 This research was supported in part by grants from the US-Israel
Binational Science Foundation.
% ===================================================================

%\newpage
% ===================================================================

%{\bf FIGURE CAPTIONS}

%\begin{figure}
%\plotone{tabdiv1.ps}
%\end{figure}
\end{document}